\documentclass[preprint,12pt]{aastex}
\usepackage{epsfig, amsmath, amsfonts}

\newcommand\bb[1] {   \mbox{\boldmath{$#1$}}  }

\newcommand\del{\bb{\nabla}}

\def\gtsima{$\; \buildrel > \over \sim \;$}
\def\gtsim{\lower.5ex\hbox{\gtsima}}

\newcommand\bcdot{\bb{\cdot}}
\newcommand\btimes{\bb{\times}}
\newcommand\vv{\bb{v}}
\newcommand\BB{\bb{B}}

\newcommand\kva{ \bb{k\cdot v_A}  }

\newcommand\dd{\partial}
\newcommand\ddt{\partial t}

\def\beq{ \begin{equation} }
\def\eeq{ \end{equation} }

\begin{document}


\title{Radiative and dynamic stability of a dilute plasma}

\author{ Steven A. Balbus\altaffilmark{1,2},
Christopher S. Reynolds\altaffilmark{3,4}}

\altaffiltext{1}{Laboratoire de Radioastronomie, \'Ecole Normale
Sup\'erieure, 24 rue Lhomond, 75231 Paris CEDEX 05, France
  \texttt{steven.balbus@lra.ens.fr}}

\altaffiltext{2}{Adjunct Professor, Department of Astronomy, University
of Virginia, Charlottesville VA 22903}

\altaffiltext{3}{Department of Astronomy, University of Maryland, College Park, MD20742}            

\altaffiltext{4}{Joint Space Science Institute (JSI), University of Maryland, College Park, MD20742}

\begin{abstract}
We analyze the linear stability of a dilute, hot
plasma, taking into account the effects of stratification and
anisotropic thermal conduction.  The work is motivated by
attempts to understand the dynamics of the
intracluster medium in galaxy clusters.   We show that
magnetic field configurations that nominally stabilize either the heat-flux driven
buoyancy instability (associated with a positive thermal gradient) or the
magnetothermal instability (negative thermal gradient)
can lead to previously unrecognized $g$-mode overstabilities.
The driving source of the overstability is either radiative
cooling (positive temperature gradient) or
the heat flux itself (negative 
temperature gradient).   While the implications of these overstabilities
have yet to be explored, we speculate that the cold fronts
observed in many relaxed galaxy clusters may be related to
their non-linear evolution.   
\end{abstract}


\maketitle

\section{Introduction}

The dynamical stability properties of a dilute plasma change dramatically
when even a weak magnetic field with no ``$J\times B$'' force is present.
This is a consequence of anisotropic transport: if the mean free
path exceeds the relevant gyroradius, heat and (angular) momentum are
transported only along magnetic lines of force (Balbus 2000, 2004).  In
particular, in a stratified atmosphere of dilute plasma, the Schwarzschild
stability criterion (i.e. specific entropy must increase upwards in a
gravitational potential) no longer governs convective stability (Balbus
2001).  Instead, the presence of any temperature gradient is potentially
convectively unstable, depending on the relative alignment of the magnetic
field and thermal gradient (Quataert 2008, hereafter Q08).  The only
configurations that are nominally stable are i) an upwardly decreasing
temperature profile whose gradient is precisely aligned with the magnetic
field, and ii) a upwardly increasing temperature profile whose gradient
is precisely orthogonal to the magnetic field.  Any other configuration
is unstable.  In this {\it Letter,} we show that even these ``stable''
configurations are in fact subject to overstability.  Interestingly, the
non-linear evolution of the thermal gradient {\it instabilities} tends
to drive the system towards these {\it overstable} states.  In essence,
a dilute stratified plasma that is hot at its core wants to let the heat
escape; a cold-core system wants to self-insulate itself.  The hot core
instability is known as the magnetothermal instability (MTI).  The cold
core instability is known as the heat flux buoyancy instability (HBI),
because the free energy source is the thermal flux itself.

Both the MTI and HBI may be of considerable importance for understanding
the behavior of the intracluster medium (ICM) in clusters of galaxies.
The outer regions of ICM atmospheres have a negative temperature gradient
and MTI-driven turbulence may be important for energy and metal transport
(Parrish, Stone \& Lemaster 2008).  Furthermore, in clusters with
cooling cores, radiative losses lead to the formation of a positive
temperature gradient in the central regions of the ICM: a cool ICM core
is embedded in a hotter ICM atmosphere (e.g., review by Peterson \&
Fabian 2006).   A longstanding problem has been to understand the thermal
stability of the cool inner regions of ICM atmospheres.  In particular,
why do radiative losses not cause run away cooling in these regions?
The standard paradigm is that regulated energy injection into the ICM
by a central active galactic nucleus (AGN) is a stabilizing influence.
However, the finding of Q08 the positive temperature gradient within the
ICM of cooling core clusters could be convectively unstable suggests that
the destabilzing role played by thermal conduction merits closer scrutiny
in this context.

A major difficulty with invoking thermal conduction to stave off
radiative core collapse of the ICM (e.g., Bertschinger \& Meiksen 1986,
Zakamska \& Narayan 2003) has been the fine tuning required (Conroy \&
Ostriker 2008).  A little too much thermal conduction evaporates the cool
core; too little allows the collapse to proceed essentially unimpeded.
Balbus \& Reynolds (2008) hypothesized that the combination of thermal
and radiative losses may be self-regulating.  In this view, suppression
of the thermal conduction is not complete.  Instead, through field
line tangling and a possible reverse HBI convective thermal flux, the
radiative core is able to draw in the requiste heat flux for marginal
stability.   Global MHD models of cluster cores have recently tested
these ideas (Bogdanovic et al. 2009; Parrish, Quataert \& Sharma, 2009,
2010; Ruszkowski \& Oh 2010).  These simulations show that in idealized
static atmospheres, while there is significant heat transport into the
cooling core, after several cooling times the non-linear development
of the HBI wraps the magnetic field onto shells, insulating the core
from further conduction and fostering thermal collapse.  Models with
sufficiently strong forced turbulence (perhaps the effects of an AGN,
or stirring by galaxy motions and/or sub-cluster mergers) can prevent
the field-line wrapping and stabilize the cluster.  However, such models
predict temperature decrements in the core of at most a factor of two;
real cooling core clusters often have much steeper temperature gradients.
Moreover, the forced turbulence is assumed to be volume-filling, the
justification for which is unclear.

In this {\it Letter}, we return to the fundamental question of the
linear stability of an atmosphere of dilute weakly magnetized plasma
in the presence of a background temperature gradient.   We generalize
the treatment of Q08 to include optically--thin radiative losses.  Our
principal result is the finding of new overstabilities of dynamical waves.
More precisely, we find that nominally stable configurations resulting
from the non-linear evolution of the HBI (i.e., temperature increasing
upwards and magnetic field essentially horizontal) generate overstable
$g$-modes via radiative losses.  Nominally stable configurations resulting
from the non-linear evolution of the MTI (i.e., temperature decreasing
upwards and magnetic field essentially vertical) generate overstable
$g$-modes via anisotropic thermal conduction.  In addition to furnishing
a more complete formal picture of the stability properties of dilute
plasma atmospheres, these findings may have significant implications
for the physical behavior of the ICM, and should guide future simulations.

In the next section, we present the calculation in detail, and in the
final section of this {\it Letter,} we conclude with a brief discussion
of the implications of our findings.

\section{Analysis}

We use the standard equations of MHD, with the entropy equation agumented 
with anisotropic thermal
conduction along magnetic field lines (Braginskii 1965) and radiative
losses (e.g. Field 1965).  The mass, momentum, induction, and entropy
equations are respectively
\beq
{\dd\rho\over\dd t} +\del\bcdot(\rho\bb{v})=0,
\eeq
\beq
\rho {D\bb{v}\over Dt}=
{ (\del\btimes\BB)\btimes\BB\over4\pi} -\del P +\rho
\bb{g},
\eeq
\beq
{\dd\BB\over\ddt} = \del\btimes (\vv\btimes\BB),
\eeq
\beq\label{ent}
{D\ln P\rho^{-\gamma}\over Dt} = - {\gamma -1\over P}
\left[ \del\bcdot\bb{Q} +\rho\cal{L}\right],
\eeq
where $\rho$ is the mass density, $\vv$ is the fluid velocity,
$\BB$ is the magnetic field, $\bb{g}$ is the gravitational
acceleration, $P$ is the gas pressure, $\gamma$ is the adiabatic
index (5/3 for monotonic gas),
$\bb{Q}$ is the heat flux, and ${\cal L}$ is the radiative
energy loss per unit mass of fluid, whose form we will leave 
unspecified.  For thermal bremsstrahlung, a reasonable
approximation is
\beq
\rho{\cal L} \simeq 2\times 10^{-27} n_e^2 T^{1/2}
{\rm ergs\ } {\rm cm}^{-3}\ {\rm s}^{-1},
\eeq
where $n_e$ is the electron number density.
$D/Dt$ is the 
Lagrangian derivative, $\dd/\ddt +\vv\bcdot\del$.

To define the heat flux $\bb{Q}$, let $\bb{b}$ be a unit vector in 
the direction of the magnetic field.  Then (Balbus 2001):
\beq
\bb{Q} = - \chi \bb{b}(\bb{b}\bcdot\del)T,
\eeq
where $T$ is the kinetic gas temperature and $\chi$ is the thermal
conductivity (Spitzer 1962):
\beq
\chi \simeq 6\times 10^{-7} T^{5/2} {\rm ergs\ }{\rm cm}^{-1}
\ {\rm s}^{-1}\ {\rm K}^{-1}.
\eeq
Finally, we follow Q08 and use the notation
\beq
\kappa\equiv \chi T/P
\eeq
for the thermal diffusion coefficient.  

\subsection{Equilibrium Background}

We consider a gas stratified in the vertical $z$ direction,
with temperature profile $T(z)$.  The gas is not self-gravitating,
so that $\bb{g}$ is a specified function of position.  We assume
a highly sub-thermal magnetic field.  Thus, in equilibrium,
the gas is in hydrostatic balance,
\beq
\frac{dP}{dz}=-\rho g.
\eeq
The magnetic field is uniform with $x$ and $z$ components
$B_x$ and $B_z$,  
(in this way defining the $x$ axis), and unit vectors
$b_x=B_x/B$, $b_z=B_z/B$.  In equilibrium, there is
a thermal balance between conductive heating 
and radiative losses,
\beq
-\del\bcdot\bb{Q}\equiv  {d^2(b_z^2\chi)T\over dz^2} = \rho{\cal L}.
\label{eqn:therm_eqm}
\eeq

\subsection{Local WKB Perturbations}

As in Q08, we consider plane wave disturbances of the form $\exp(\sigma
t+i\bb{k\cdot r})$ where the wavenumber $\bb{k}$ has Cartesian components
($k_x$, $k_y$, $k_z$), and $\bb{r}$ is the position vector.  We differ in
notation from Q08 by using $\sigma$, a formal growth rate, rather than
$\omega$, an angular frequency.   This ensures that all coefficients in
the final dispersion relation are real.  We work in the WKB ($kr \gg 1$)
and Boussinesq limits (Q08).

We next consider the linearized equations when 
perturbations $\delta\rho$, $\delta \vv$, etc., are applied to the
equilibrium state.  The heart of the problem is the entropy equation, 
so let us begin here.  The linearized form of equation (\ref{ent}) is
\beq\label{linent}
-\gamma\sigma {\delta\rho\over \rho} + \delta v_z{d\ln P\rho^{-\gamma}
\over dz} = (\gamma-1)\left[ -{\del\bcdot\delta\bb{Q}\over P}
-\Theta_{T|P}\ \delta T\right],
\eeq
where
\beq
\Theta \equiv \rho{\cal L}/P,
\eeq
and
\beq
\Theta_{T|P}\equiv \left[{\dd\Theta \over \dd T}\right]_P,
\eeq
that is, the derivative of $\Theta$ with respect to $T$ with $P$
held constant.  We have used the Boussinesq approximation in ignoring
all terms proportional to $\delta P$ in equation (\ref{linent}).  In
the process, we
have implicitly regarded $\Theta$ as a function of $T$ and $P$ (rather
than the more customary but less convenient $T$ and $\rho$
dependence).  The remaining linearized equations,
\beq
\bb{k}\bcdot\vv=0,
\eeq
\beq
\sigma\delta\vv= {\delta\rho\over\rho^2}{dP\over dz} -i\bb{k}
\left( {\delta P\over\rho} + {\BB\bcdot\delta\BB\over 4\pi\rho}\right)
+{i(\bb{k\cdot B})\delta\BB\over 4\pi\rho},
\eeq
\beq
\sigma\delta\BB = i(\bb{k\cdot B})\delta\vv,
\eeq
are, apart from notational convention, identical to Q08.  The
entire system of equations differs from Q08 only by the $\Theta$ term.
The resulting dispersion relation is                     
$$
\left(
\sigma + {\gamma-1\over\gamma}T\Theta_{T|P} +{\cal C}
\right)
\left(
\sigma^2+(\bb{k\cdot v_A})^2\right)\qquad\qquad\ \ \ \ \ \
$$
\beq\label{disp}
+{\sigma k_\perp^2 N^2\over k^2}
+{\cal CK} {g\over k^2}{d\ln T\over dz}=0,
\eeq
where
\beq
N^2 = - {1\over \rho \gamma}{dP\over dz}{d\ln P\rho^{-\gamma}\over dz}=
g{d\ln P^{(1-\gamma)/\gamma}T\over dz},
\eeq
\beq
{\cal C}=\left(\gamma-1\over\gamma\right)\kappa (\bb{k\bcdot b})^2,
\eeq
\beq
{\cal K}= (1-2b_z^2)k_\perp^2+2b_xb_zk_xk_z.
\eeq
This corresponds to equation (13) of Q08, except, as noted,
for the 
single appearance of the radiative $\Theta_{T|P}$ term.  
(A less general version of this result was also presented
in Balbus \& Reynolds [2008].)  The dispersion characterizes the linear
response of a magnetized, thermally conducting radiative
dilute plasma to incompressible disturbances.

\subsection {Stability}

\subsubsection{Recovery of the Conductive Field criterion}

Expanding the dispersion relation (\ref{disp}) leads to
\beq
\sigma^3 + a_1\sigma^2 +a_2\sigma + a_3 = 0,
\eeq
where
\beq
a_1 = \left(\gamma-1\over \gamma\right) T\Theta_{T|P} +{\cal C},
\eeq
\beq
a_2 = {k_\perp^2\over k^2}N^2 +(\bb{k\cdot v_A})^2,
\eeq
\beq
a_3={\cal CK} {g\over k^2}{d\ln T\over dz}+(\bb{k\cdot v_A})^2 a_1.
\eeq
There are stable solutions to this dispersion relation if and only
if the following three criteria are met:
\beq
a_1>0, \quad a_3>0, \quad a_1a_2 >a_3.
\eeq
This follows from a Routh-Hurwitz analysis, but can be seen more easily
by inspection: the first two are in fact elementary, while the third
follows from self-consistently
demanding purely imaginary solutions to the cubic equation and
then investigating their behavior for infinitesimal real parts.  

The physical interpretation of $a_1>0$, or
\beq\label{26}
T\Theta_{T|P} + (\bb{k\bcdot b})^2\kappa  >0,
\eeq
is the magnetized conduction variation of the classical thermal
instability
criterion (Field 1965).   Only the component of $k$ along the field lines
enters into the conduction term.

\subsubsection{Recovery of the HBI and MTI}

We next consider the physical interpretation of $a_3>0$, or
\beq\label{27}
{\cal CK} {g\over k^2}{d\ln T\over dz}+(\bb{k\cdot v_A})^2 a_1>0.
\eeq
In essence, this is the HBI/MTI criterion of Q08, but further (de)stabilized when
the flow is (un)stable by the isobaric Field criterion.  This is a true
instability if $a_3$ is negative, with $\sigma=-a_3/a_2$ in the
limit of large $a_2>0$.  

Equation (\ref{27}) shows that thermal instability and the HBI/MTI are 
intimately linked.   To be definite, consider the behavior of the HBI.
The discussion of Q08 explains how the distortion of
field lines leads to conductive cooling of a downwardly displaced fluid
element (and vice-versa for an upwardly displaced element).  It is
this cooling that causes the convection associated with the HBI.  With
radiative losses present, the cooling is enhanced on a downward displacement,
and relative heating is present on an upward displacement. 
In fact, we may imagine now slowly turning on the magnetic
field from dynamically weak to strongly dominant.  Then, the exponentially
growing instability transforms from the Q08 HBI to the classical
(nonoscillatory) thermal instability.  The role of the magnetic field
in mediating this transition is crucial.  

\subsubsection{Destabilization of wave modes by a positive thermal
gradient}

We now return to the third criterion, $a_1a_2>a_3$.
With $a_3>0$, this criterion
is a more stringent stability criterion than the first ($a_1>0$), 
and hence replaces it.  

When $a_2$ is large and positive [e.g. either $N^2$ or
$(\kva)^2$ is dominant], 
the unstable roots depending on $a_1$ will be approximately:
\beq\label{XX}
\sigma = \pm ia_2^{1/2} +(a_3-a_1a_2)/2a_2
\eeq
On the other hand, at large wavenumbers, we may have $a_1$ and $a_3$ 
as the dominant terms.  If $a_3$ and $a_1$ are both positive (or both
negative), then
the wavelike solutions will be
\beq
\sigma = \pm i (a_3/a_1)^{1/2} + (a_3-a_1a_2)/2a_1^2
\eeq
In either case above, the combination $a_3-a_1a_2$ determines the stability
of the mode.  

After a cancellation of the magnetic tension terms, the condition
$a_1a_2-a_3>0$ becomes
\beq\label{28}
a_1 {k_\perp^2\over k^2} N^2 - {\cal CK}{g\over k^2}{d\ln T\over dz} >0.
\eeq
Consider the limit $b_z\ll 1$, which is HBI stable ($a_3>0$)
for all but nearly axial
wavenumbers, whose growth times then become very long.   (We cannot take $b_z=0$
exactly, since that would preclude a static radiative equilibrium state.
For $b_z$ finite, equation (\ref{eqn:therm_eqm}) shows that 
the equilibrium $dT/dz$ scales as $b_z^{-1}$.)
Then, ${\cal K} =k_\perp^2$, and our inequality reduces to                  
\beq\label{29}
{\gamma-1\over \gamma}\  T\Theta_{T|P}\ N^2 +{\cal C}
\left(N^2 -g{d\ln T\over dz}\right)>0.
\eeq
But
\beq
N^2 -g{d\ln T\over dz}
={\gamma-1\over\gamma}{1\over P\rho}\left(dP\over dz\right)^2,
\eeq
and assuming that $N^2>0$, the inequality may be yet further reduced:
\beq
T\Theta_{T|P} +{ {\cal C}\over \rho P N^2}\left(dP\over dz\right)^2 >0.
\eeq
Finally, substituting for ${\cal C}$ and $N^2$ and simplifying, 
our condition becomes
\beq
T\Theta_{T|P} + \kappa (\bb{k\cdot b})^2 {\cal R}>0,
\eeq
where ${\cal R}$ is the reduction factor
\beq
{\cal R} = \left( 1 + \left| {\gamma\over \gamma -1}{d\ln T\over d\ln P}
\right|\right)^{-1},
\eeq
which bears direct comparison with equation (\ref{26}).  Here, thermal
conduction once again stabilizes radiative losses, but the HBI terms,
when they are in a {\em stable} configuration relative to the convective
processes discussed by Q08, actively {\em destabilize} wave-like modes
by reducing the supression of thermal conduction.
In regions of sharp temperature gradients, the effective reduction factor for
conductive stabilization can be large.  Indeed, in the chosen limit
$b_z \ll 1$, we have 
${\cal R}\sim O(b_z)$.  Note that wavenumbers with vanishing
$\bb{k\cdot b}$ are unaffected by conduction, and have an
effective reduction factor of zero.  In our example, these are horizontal
fluid displacements along the magnetic field lines.

\subsubsection{Destabilization of wave modes by a negative thermal
gradient}

Consider next the case $b_z=1$, which would be HBI-unstable in the 
case of an increasing outward temperature profile.  But let us now assume
that the temperature decreases outwards.  This configuration is
HBI stable.  With $b_z=1$, if we restricted
ourselves only to the first two stability criteria, we would conclude that
this configuration is also MTI-stable.  In fact, if the third
stability criterion is imposed, this configuration is subject to an
interesting and powerful overstability, driven by anisotropic
thermal conduction, as we now show.   

With $b_z=1$ we have ${\cal K} = - k_\perp^2$ and our third criterion inequality (\ref{28}) 
becomes
\beq\label{31}
{\gamma-1\over \gamma}\  T\Theta_{T|P}\ N^2 +{\cal C}
\left(N^2 +g{d\ln T\over dz}\right)>0.
\eeq
Once again, the thermal conduction is affected by a ``reduction
factor,'' though here the reduction factor ${\cal R'}$ 
actively destabilizes rather than merely supresses dissipative
destabilization.  The above inequality may be written
\beq
T\Theta_{T|P} + \kappa (\bb{k\cdot b})^2 {\cal R'}>0,
\eeq
where
\beq
{\cal R'} = 1 - \left[ {\gamma-1\over\gamma}{d\ln P\over d\ln T}
-1 \right]^{-1}
\eeq
The term inside the square brackets
must always be positive if $N^2>0$, but if
\beq
1<{\gamma-1\over\gamma}{d\ln P\over d\ln T}<2
\eeq
then ${\cal R'}<0$ and buoyant modes are overstable, {\em even
if there is no radiative loss term.}

\subsubsection{Summary}

The MTI and HBI are evansecent instabilities present in dilute plasmas
when anisotropic heat flux is included in the physics.  The MTI is present
when the thermal gradient decreases outward and the field lines are
insulating in the equilibrium configuration.  When the field lines open,
the MTI is stabilized.  The HBI is present when the thermal gradient increases 
outward and the field lines are open so that a heat flux is present in the
equilibrium configuration.  The action of the HBI is to close the field lines,
which stabilizes the system.

We have found that the stable ``end states'' of these instabilities are
subject to further overstabilities.  In the case of the HBI, which is
relevant for the coooling flow cores, a thermally unstable radiative loss
function and closed fields lines together manifest as over stable 
buoyant oscillations.  In the case of
the MTI, a sufficiently steep (but classically convectively stable) outwardly
decreasing thermal gradient produces overstable buoyant waves when the magnetic
field lines are open and conducting heat.  

The overstabilities nominally depend on radiative losses, but their
effect should be thought of as dynamical: these are classical g-waves
that in principle could be driven to finite amplitudes on thermal
time scales (either radiative or conductive).   Whether they are best
thought of a local WKB waves, global modes, or both is not yet clear,
and awaits numerical investigation.

\section {Discussion and Conclusions}

The implications of the Q08 finding that generic cluster (or elliptical
galaxy) cooling flows are convectively unstable have yet to be grasped. A
more complete linear theory is clearly a starting point.   Here, we have
generalized the linear theory of such systems to include the effects of
both anisotropic thermal conduction and optically-thin radiative losses.

To recap, strict stability requires three criteria to be satisfied.
The first amounts to  the classical Field criterion for thermal
instability in the presence of anisotropic conduction,
\begin{equation}
a_1\equiv T\Theta_{T|P} + \kappa(\bb{k\bcdot b})^2  >0\hspace{1cm}{\rm (Stability)}.
\label{eqn:field}
\end{equation}
The second criterion gives the MTI or HBI stability conditions depending
upon the orientation of the magnetic field (via ${\cal K}$) and the
temperature gradient,
\begin{equation}
{\cal CK} {g\over k^2}{d\ln T\over dz}+(\bb{k\cdot v_A})^2 a_1>0\hspace{1cm}{\rm (Stability)}.
\label{eqn:hbi}
\end{equation}
The third criterion has not, to our knowledge, been
recognized previously.  For the two limiting cases considered
in this work, it takes the form
\beq
T\Theta_{T|P} + \kappa(\bb{k\bcdot b})^2 {\cal R} >0\hspace{1cm}{{\rm (Stability\ }; b_z\approx 0; dT/dr>0)},
\label{eqn:overstab1}
\eeq
\beq
T\Theta_{T|P} + \kappa(\bb{k\bcdot b})^2 {\cal R}^\prime >0\hspace{1cm}{{\rm (Stability\ }; b_z\approx 1; dT/dr<0)},
\label{eqn:overstab2}
\eeq
where $0<{\cal R}<1$ and $-\infty<{\cal R}^\prime<1$.  Even once the
HBI (MTI) has been stabilized by the formation of horizontal (vertical)
magnetic fields during their non-linear evolution, the third criterion can
be violated in some range of wavenumbers leading to overstable $g$-modes.

The mechanism of the overstability differs depending upon the setting.
The overstability that would naturally follow HBI-driven evolution
($dT/dz>0$ and essentially horizontal field) is driven by the
presence of Field-unstable radiative cooling.   On the other hand, the
overstability that would naturally follow MTI-driven evolution ($dT/dz<0$
and essentially vertical field) is driven by the anisotropic heat flux,
and is present even in the absence of radiative losses.

\begin{figure}[t]
\centerline{
\psfig{figure=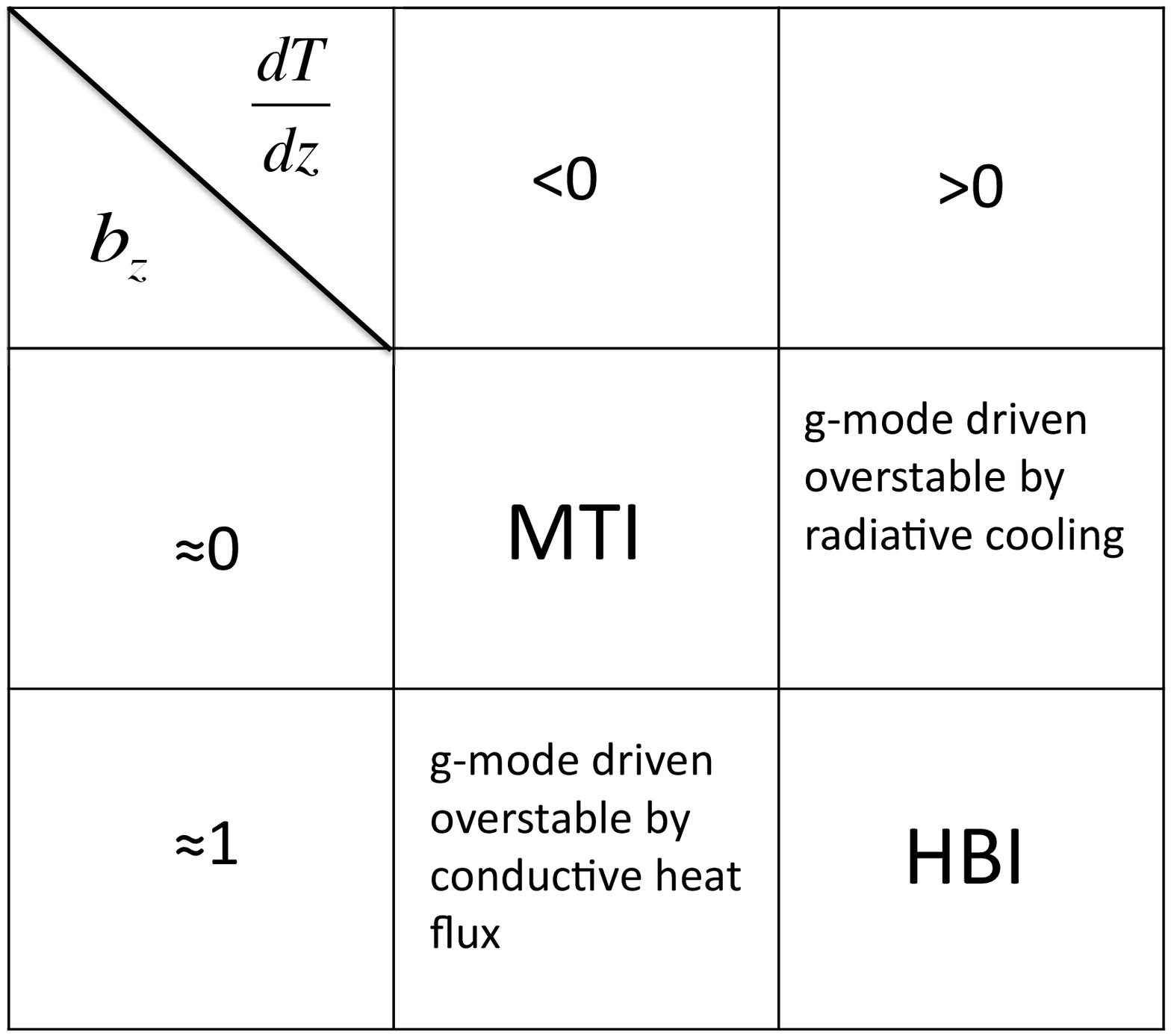,width=0.6\textwidth,angle=270}
}
\caption{Schematic map of the instabilities and overstabilities
discussed in this work.}
\end{figure}

Figure~1 indicates schematically the set of instabilities and 
overstabilities discussed in this paper.
At a formal level, it is remarkable that the conductive heat flux
can drive the atmosphere away from stability irrespective of the 
sign of the temperature gradient: only very shallow negative 
temperature gradients 
[such that $d\ln P/d\ln T>2\gamma/(\gamma-1)$] are completely
stable, and even here the overstability can be reinstated if
Field-unstable radiative losses are present.    

The astrophysical implications of these overstabilities have yet to be
determined, and we must await simulations in order to assess their global
character as well as their non-linear behaviour\footnote{To date, published 
local studies of the HBI have excluded radiative cooling,
and so are not be subject to this overstability.  The overstabilities
should, however, be present in published global studies of 
cooling cores (Bogdanovic et al. 2009; Parrish et al. 2009).   
But these simulated cores undergo a thermal runaway
after several cooling times, and the overstabilities may be buried in
the complex background dynamics, and very difficult to extract.}  
Here, we limit ourselves to some brief remarks and speculation.
Probably the most important implication of these overstabilities for
cooling core clusters is that, provided that radiative losses maintain a
temperature gradient, the ICM may never be able to achieve a state that is
stable to the HBI and the $g$-mode overstability.  This statement holds
true even if some unspecified heat source is fending off core collapse,
provided that the {\it net} cooling function is still Field-unstable.
The $g$-mode overstability grows on the cooling timescale and can be
easily seeded by perturbations resulting from AGN activity, galaxy
motions, sub-cluster mergers or indeed a preceeding phase of HBI-driven
turbulence (Ruszkowski \& Oh 2010).   The $g$-mode overstability may still
be relevant even when forced turbulence erases any obvious manifestations
of the HBI.   It is interesting to speculate that the ``cold fronts"
seen in the ICM cores of many relaxed clusters (Markevitch et al. 2000;
Ascasibar \& Markevitch 2006) may be related to the non-linear development
of these radiatively-driven $g$-mode overstabilities.

At the purely formal level, understanding the development of thermal 
instability in a dilute plasma (Field 1965) has turned into a 
four decade struggle.  And one speaks here of {\em linear}
stability!  It is a problem of some subtlety.  With the advent
of the HBI and our deepened understanding of the role of anisotropic
conduction, we may at last have the true essence of the problem.
There may, on the other hand, be more to come.

\section*{Acknowledgements}
CSR thanks the Physics Department of the \'Ecole Normale
Sup\'erieure de Paris for its hospitality and support of a one
month visit in June 2010 during which some of this work was
conducted.  This work is supported by the NSF under
grant AST 09-08212.

\label{lastpage}

\end{document}